\documentclass[aps, prb, reprint, showpacs]{revtex4-1} 

\usepackage{amsmath}
\usepackage{graphicx}
\usepackage{amsmath}
\usepackage{amssymb}
\usepackage{url}
\usepackage{bm}

\begin{document}

\title{Radiative coupling of quantum dots in photonic crystal structures}

\author{Momchil Minkov and Vincenzo Savona}

\affiliation{Institute of Theoretical Physics, Ecole Polytechnique F\'{e}d\'{e}rale de Lausanne EPFL, CH-1015 Lausanne, Switzerland}

\date{\today}

\begin{abstract}
We derive a general formalism to model the polariton states resulting from the radiation-matter interaction between an arbitrary number of excitonic transitions in semiconductor quantum dots and photon modes in a photonic crystal structure in which the quantum dots are embedded. The Maxwell equations, including the linear nonlocal susceptibility of the exciton transitions in the quantum dots, are cast into an eigenvalue problem, which can be applied to any structure whose photon modes can be computed with reliable accuracy, and in addition naturally allows for disorder effects to be taken into account. We compute realistic photon modes using Bloch-mode expansion. As example systems, we study typical InGaAs quantum dots in a GaAs photonic crystal structures -- an $Ln$ cavity or a $\mathit{W1}$ waveguide. For a single dot, we reproduce known analytical results, while for the two dot case, we study the radiative excitation transfer mechanism and characterize its strength, the dependence on the detuning between quantum dot and photon modes, and the dependence on inter-dot distance. We find in particular that the inter-dot radiative coupling strength can reach $100 \mu{eV}$ in a short cavity, and its decay with distance in longer cavities and waveguides is determined by the group velocity of the exchanged photons and their radiative lifetime. We also show that, for an $Ln$ cavity of increasing length, the radiative excitation transfer mechanism is subject to a crossover from a regime where a single photon mode is dominating, to a multi-mode regime -- occurring around $\mathit{n}$ = 150 for the system under study. 
\end{abstract}


\maketitle

\section{Introduction} 

In the last two decades, the design and implementation of devices for quantum information processing has been a major goal of condensed matter physics. An essential requirement of the quantum information paradigm is the possibility for two \textit{qubits} to interact coherently in a controlled fashion, in order to achieve controlled gate operations. This must in principle be possible for each arbitrarily chosen pair of qubits in the system. Most of the technologies, however, employ qubits which are at all times spatially separated and do not interact directly \cite{DiCarlo2009,Riebe2004,Weperen2011}. The interaction can then be achieved by means of a \textit{quantum bus}, namely a spatially extended degree of freedom interacting with all localized qubits. In a more general picture, these spatially extended degrees of freedom might even form a quantum network connecting distant quantum information systems.\cite{Kimble2008} A quantum bus can be of several kinds -- two common examples being phonons in chains of trapped ions\cite{Cirac_1995} and microwave photons in superconducting circuits.\cite{Sillanpaeae_2007, Majer_2007}. Photons are the most natural choice in a solid-state system, given their low decoherence rate, high velocity, and the recent advances in the on-chip photonic technology, especially in the photonic crystal (PHC) domain \cite{Noda2007, Notomi2010}. There, extremely high-$\mathcal{Q}$ cavities with modal volumes of the order of $(\lambda/n)^3$ have been fabricated both in silicon\cite{Kuramochi_2006, Taguchi_2011} and in GaAs\cite{Combrie_2008}, as well as waveguides allowing for low-loss, long-range photon transfer with a controllable group velocity.\cite{Baba_2008} The advance in PHC technology opened the way to several experimental breakthroughs, including low power all-optical switching,\cite{Husko_2009, Nozaki_2010} and the dynamic control of the strong coupling between two distant cavities\cite{Sato_2012} -- highlighting the extreme level of control of light which is currently achievable.

Semiconductor quantum dots (QDs) have long been considered as viable qubit candidates,\cite{Loss_1998} as they naturally fulfill the criteria of scalability and integrability required in a quantum information technology. Facing the remarkable advance made in the system of spin qubits in lateral QDs \cite{Weperen2011} -- where electron spins are controlled electronically with ohmic contacts -- optical excitations in self-organized QDs have only recently caught up in the race towards controlled quantum operations. On one hand, in fact, full single-qubit optical control has been successfully demonstrated.\cite{Patton_2005, Press2008, Berezovsky_2008, Greilich_2009, Greilich_2011, Poem_2011, Muller_2012, Godden_2012, Kodriano_2012} On the other, integrating QDs in photonic structures has made significant progress, and both single-dot Purcell enhancement in cavities\cite{Gerard_1998, Solomon_2001, Happ_2002, Ramon_2006, Munsch_2009} and waveguides\cite{Viasnoff-Schwoob_2005, Lund-Hansen_2008, Thyrrestrup_2010, Schwagmann_2011, Hoang_2012, Laucht_2012, Laucht_2012a}, and strong coupling to a cavity mode\cite{Yoshie_2004, Reithmaier_2004, Peter_2005} have been demonstrated. Single-dot coupling to light modes is in itself important for practical applications, as suggested by the possibility of non-classical light generation,\cite{Faraon_2008, Kasprzak_2010,Dousse2010, Reinhard_2012} or single-photon optical switching.\cite{Volz_2012} Beyond that, short-distance coupling in quantum dot ``molecules'' has been demonstrated,\cite{Bayer_2001, Borri_2003, Gerardot_2005, Krenner_2005, Stinaff_2006,Robledo2008} where however the coupling is enforced by the direct overlap of the QD wave-functions and/or the electrostatic F\"{o}rster dipole-dipole interaction,\cite{Govorov_2003, Govorov_2005} rather than by any long-distance mechanism. Altogether, these advances suggest that the field has reached the milestone, following which the process of \textit{long-distance}, photon-mediated interaction between two or more quantum dots should also be addressed. It has been shown that the light-matter interaction between a QD and the electromagnetic modes of a non-structured photonic environment is very weak,\cite{Parascandolo_2005, Scheibner_2007, Tarel_2008, Kasprzak_2011} thus photonic structures are needed in order to tailor the density of optical modes and thus enhance radiative coupling between spatially separated quantum dots. Indeed, short-range radiative coupling has already been achieved in several experiments involving small optical cavities, where strong coupling of two quantum dots to the same cavity mode was detected,\cite{Reitzenstein_2006, Laucht_2010, Kim_2011} and, most recently, its coherent nature was demonstrated.\cite{Albert_2012} A photonic structure has also brought the experimental demonstration of long-distance transfer of photons emitted by an embedded QD.\cite{Englund2007}

On the theoretical side, specific aspects of structures with one or more quantum dots in a photonic environment have been studied. These include the strong coupling regime and emission spectrum of one\cite{Andreani_1999, Khitrova_2006, Milde2008, Tarel_2010,Hughes2011} or more\cite{Kessler_2008, Laussy_2011, Auffeves_2011} dots in a microcavity, as well as the possibility of performing cavity-mediated qubit operations through coherent excitation exchange in such a system.\cite{Agarwal_1998, Imamoglu_1999, Piermarocchi_2002, Quinteiro_2006, Xu_2011} In addition, the spontaneous emission enhancement of one dot coupled to a single waveguide mode has been estimated\cite{Hughes_2004, Rao_2007, Lecamp_2007}, and non-trivial dynamics of single-dot cavity-QED in presence of coupling to a second, distant cavity, have been predicted\cite{Hughes_2007}. There are, however, only a few studies of the dot-dot interaction at a mesoscopic (i.e. more than one wavelength) inter-dot distance -- which is a main focus of this work. Most notably, the possibility to generate entangled states between distant QDs in a coupled-cavity system was recently demonstrated,\cite{Yao_2009} as well as the non-trivial decay dynamics\footnote{Here we prefer not to use the term ``superradiant'', in order to avoid ambiguity with the concept of Dicke superradiance, which has a radically different physical nature.} of two distant dots in a photonic crystallite.\cite{Kristensen_2011} However, a general formalism accounting for an arbitrary number of quantum dots coupled to arbitrarily many photonic modes is still lacking, and in particular, the distance dependence of the radiative interaction, the influence of fabrication disorder, and the competition between excitation transfer at-a-distance and radiation  losses still remain open questions. To address those, a microscopic description of light-matter coupling with a realistic description of the photonic modes is needed.

In this paper, we lay down the semi-classical linear response theory for a system of $N$ distinct, spatially localized excitonic transitions in QDs, coupled to $M$ photonic modes of an arbitrary photonic structure. In particular, we frame the underlying Maxwell equations into an eigenvalue problem, describing the polariton modes of the system in analogy with the polariton formalisms for a bulk semiconductor\cite{Hopfield_1958}, for quantum wells\cite{Tassone_1992}, and for QDs in an unstructured photonic environment.\cite{Parascandolo_2005,Tarel_2008} For the computation of the photonic modes, the Bloch-mode expansion method is employed,\cite{Savona_2011} although any other method which provides reliable field profiles (e.g., finite-element
method (FEM), finite-difference time-domain (FDTD)) can also be used. Even though the modeling of radiative effects in presence of fabrication disorder lies beyond the scope of the present work, the Bloch-mode expansion is particularly well-suited for treating large, disordered photonic structures\cite{Savona_2011, Minkov_2012} and was thus an obvious choice in view of a future extension to disordered PHCs. We apply the formalism to $Ln$ cavities and $W\mathit{1}$ waveguides based on a PHC slab. We show how known single-dot radiative properties -- such as the vacuum Rabi splitting in a microcavity and the Purcell enhancement and $\beta$-factor in a waveguide -- are well reproduced. The main focus of the work however is the quantitative characterization of radiative coupling between \textit{two} dots in those same structures. To this purpose, we characterize the spectra of the polariton eigenmodes, the time-evolution of a starting excitation in one of the dots, and the distance-dependence of the radiative excitation transfer in a spatially extended structure. Our simulations provide a comprehensive picture of the effective dot-dot radiative coupling, and show that, with realistic PHC and QD parameters, a sizable interaction can be expected at а mesoscopic distance. 

The work is organized as follows. In Section II we derive the main theoretical formalism, while in Section III we thoroughly discuss the values of the parameters entering the model, for realistic InAs/GaAs-based semiconductor nanostructures. Section IV contains the main results obtained from the application of the model to the study of one and two QDs embedded in $Ln$ cavities and $W\mathit{1}$ waveguides. In Section V we present our conclusions and an outlook of future work.
 
\section{Theoretical formalism}

Starting from Maxwell's equations with the assumptions of a non-magnetic medium and no free charges, the electric field in the frequency domain obeys the equation (written in Gaussian units) 

\begin{equation}
\nabla \times \nabla \times \mathbf{E}(\mathbf{r}, \omega) - \frac{\omega^2}{c^2}\left(\varepsilon(\mathbf{r}) \mathbf{E}(\mathbf{r}, \omega) + 4 \pi \mathbf{P}(\mathbf{r}, \omega) \right) = 0 \, .
 \label{starting}
\end{equation}
In particular, here the spatial dependence of the dielectric constant, $\varepsilon(\mathbf{r})$, completely characterizes the underlying photonic structure, while the optical response of the quantum dots is included in the polarization vector through a non-local susceptibility tensor\cite{Kubo_1957}, such that 

\begin{equation}
\mathbf{P}(\mathbf{r},\omega) = \int \mathrm{d}\mathbf{r}' \hat{\chi}(\mathbf{r}, \mathbf{r}', \omega) \mathbf{E}(\mathbf{r}', \omega) \, .
\end{equation} 

In what follows, we will consider the specific case of excitons originating from the heavy-hole band of a semiconductor with cubic symmetry (e.g. InAs), for which only the $\mathbf{x}-$ and $\mathbf{y}-$components of the polarization couple to the electromagnetic field according to the following susceptibility tensor\cite{Tassone_1990,Tassone_1992,Andreani_1994}

\begin{equation}
\hat{\chi}(\mathbf{r}, \mathbf{r}', \omega) = \frac{\mu_{cv}^2}{\hbar} \sum_{\alpha = 1}^N \frac{\Psi_{\alpha}^*(\mathbf{r}) \Psi_{\alpha}(\mathbf{r}')}{\omega^{\alpha} - \omega} 
\begin{pmatrix}
1 & 0 & 0 \\
0 & 1 & 0 \\
0 & 0 & 0 
 \end{pmatrix} \, . 
 \label{susc}
\end{equation}
The formalism can be easily generalized to different forms of the susceptibility tensor. Here, $\alpha$ runs over all QDs, $\mu_{cv}^2$ is the squared dipole matrix element of the inter-band optical transition, $\Psi_{\alpha}(\mathbf{r}) = \Psi_{\alpha}(\mathbf{r}_e = \mathbf{r}, \mathbf{r}_h = \mathbf{r})$, and $\Psi_{\alpha}(\mathbf{r}_e, \mathbf{r}_h)$ is the excitonic wave-function, normalized as

\begin{equation}
\int \mathrm{d} \mathbf{r}_e \int \mathrm{d} \mathbf{r}_h \vert \Psi_{\alpha}(\mathbf{r}_e, \mathbf{r}_h)\vert ^2 = 1 \, .
\label{psinorm}
\end{equation}
We denote the frequencies of the bare excitons by a \textit{superscript} $\alpha$, in order to distinguish them from the frequencies of the photonic resonances, which we will later on index with \textit{subscripts}, e.g. as $\omega_m$. Notice also that here all frequencies are assumed to be complex quantities, e.g.  $\omega^{\alpha} = \Re(\omega^{\alpha}) - i\frac{\gamma^{\alpha}}{2}$, where $\gamma^{\alpha}$ represents the overall decay rate of the exciton state, including any possible non-radiative mechanism and the rate of radiative decay into photon modes that are \textit{not} included among the $M$ modes treated exactly.

In order to turn the Maxwell equation into a self-adjoint form, we introduce the quantities \cite{Sakoda_2001} $\mathbf{Q}(\mathbf{r}, \omega) = \sqrt{\varepsilon(\mathbf{r})} \mathbf{E}(\mathbf{r}, \omega)$. Eq. (\ref{starting}) then becomes

\begin{align}
\Upsilon & \mathbf{Q}(\mathbf{r}, \omega) - \frac{\omega^2}{c^2} \mathbf{Q}(\mathbf{r}, \omega) =  \label{eqforq} \\  
&\frac{4 \pi}{\sqrt{\varepsilon(\mathbf{r})}} \frac{\omega^2}{c^2}  \int \mathrm{d}\mathbf{r}' \hat{\chi}(\mathbf{r}, \mathbf{r}', \omega) \frac{\mathbf{Q}(\mathbf{r}', \omega)}{\sqrt{\varepsilon(\mathbf{r}')}}. \nonumber
\end{align}
which is an inhomogeneous differential equation defined for the \textit{self-adjoint} differential operator

\begin{equation}
\Upsilon = \frac{1}{\sqrt{\varepsilon(\mathbf{r})}} \nabla \times \nabla \times \frac{1}{\sqrt{\varepsilon(\mathbf{r})}} \; \; .
\end{equation}
The susceptibility tensor as given in Eq. \ref{susc} decouples the $\mathbf{z}$-polarized fields. We then define the two-dimensional field $\mathbf{Q} = (Q_x, Q_y)$. We can solve the problem using a Green's function approach\cite{Martin_1998}, in which the formal solution to Eq. (\ref{eqforq}) is

\begin{align}
\mathbf{Q}(\mathbf{r}, \omega) &= \mathbf{Q}_0(\mathbf{r}, \omega) + \label{dyson}\\ 
 \frac{4\pi}{\sqrt{\varepsilon(\mathbf{r})}}\frac{\omega^2}{c^2}&\int \mathrm{d}\mathbf{r}' \int \mathrm{d}\mathbf{r}'' \hat{G}(\mathbf{r}, \mathbf{r}', \omega) \frac{\hat{\chi}(\mathbf{r}', \mathbf{r}'', \omega)}{\sqrt{\varepsilon(\mathbf{r}'')}} \mathbf{Q}(\mathbf{r}'', \omega). \nonumber
\end{align}

The Green's tensor can be expanded onto the basis of field eigenmodes using the resolvent representation, following Fredholm's theory\cite{Economou_2006}

\begin{equation}
\hat{G}(\mathbf{r}, \mathbf{r}', \omega) = \sum_{m} \frac{\mathbf{Q}_m(\mathbf{r})\otimes \mathbf{Q}_m^*(\mathbf{r}')}{\frac{\omega_m^2}{c^2} - \frac{\omega^2}{c^2}} \, , 
\label{green}
\end{equation}
where the $\mathbf{Q}_m$-s are the \textit{orthonormal} eigenfunctions of $\Upsilon$ corresponding to eigenvalues $\omega_m^2/c^2$, and $\otimes$ is an outer product defined as

\begin{equation}
\mathbf{A}\otimes\mathbf{B} = 
\begin{pmatrix}
A_x B_x & A_x B_y \\
A_y B_x & A_y B_y
\end{pmatrix} \, .
\end{equation}
The sum in Eq. (\ref{green}) runs in principle over the infinite set of eigenmodes. In most situations of interest, however, this sum is dominated by the resonant modes of the photonic crystal that are closest to the frequency range characterizing the excitonic transitions. In addition, in all structures of interest (e.g. a PHC\cite{Yoshie_2004, Badolato_2005}, pillar cavity \cite{Reithmaier_2004} or a microdisc \cite{Peter_2005}), the dots are typically embedded within the dielectric medium, i.e. their wave-functions are non-negligible only in a region where $\varepsilon(\mathbf{r}) = \varepsilon_{\infty}$, the permittivity of the semiconductor. Thus, as the $\mathbf{r}$-dependence of all quantities will eventually enter through overlap integrals with the QD wave-functions, in eq. (\ref{dyson}) we can safely substitute  $\sqrt{\varepsilon(\mathbf{r})} = \sqrt{\varepsilon(\mathbf{r}'')} = \sqrt{\varepsilon_{\infty}}$. Finally, in typical situations, all QD transition frequencies lie within a small range originating from the inhomogeneous distribution of QD sizes. A very good approximation consists then in replacing the $\omega$ on the r.h.s. of (\ref{dyson}), as well as the $(\omega_m+\omega)/2$ obtained by factoring the denominator in (\ref{green}), with an average exciton transition frequency $\omega_0$. In order to compute the complex frequency poles, corresponding to the resonances of the coupled system, we consider the homogeneous problem associated with Eq. (\ref{dyson}). Then, by defining 

\begin{equation}
\mathbf{Q}^{\alpha}(\omega) = \int \mathrm{d}\mathbf{r} \Psi_{\alpha}(\mathbf{r}) \mathbf{Q}(\mathbf{r}, \omega) \, ,
\label{overlap}
\end{equation}
we obtain

\begin{equation*}
\mathbf{Q}(\mathbf{r}, \omega) = \frac{2\pi \omega_0}{\varepsilon_{\infty}}\frac{\mu_{cv}^2}{\hbar} \sum_{\alpha=1}^N \sum_{m = 1}^{M}\frac{\mathbf{Q}_m(\mathbf{r})\otimes \mathbf{Q}_m^{\alpha*}}{(\omega_n- \omega)(\omega^{\alpha} - \omega)} \mathbf{Q}^{\alpha}(\omega).
\end{equation*}
By integrating Eq. \ref{dyson} with $\int \mathrm{d}\mathbf{r} \Psi_{\beta}(\mathbf{r})$ and defining additionally $\tilde{\mathbf{Q}}^{\alpha}(\omega) = \mathbf{Q}^{\alpha}(\omega)/(\omega^{\alpha} - \omega)$, we finally obtain a set of equations (labeled by $\beta$) for the complex frequency poles

\begin{equation}
(\omega^{\beta} - \omega)\tilde{\mathbf{Q}}^{\beta}(\omega) = \frac{2\pi \omega_0}{\varepsilon_{\infty}}\frac{\mu_{cv}^2}{\hbar} \sum_{\alpha=1}^N \sum_{m = 1}^{M}\frac{\mathbf{Q}_m^{\beta}\otimes \mathbf{Q}_m^{\alpha*}}{(\omega_n- \omega)} \tilde{\mathbf{Q}}^{\alpha}(\omega) \, .
\label{alphbet}
\end{equation}

We now define the quantities 

\begin{equation}
\mathbf{g}_m^{\alpha} = ({g}_{m, x}^{\alpha}, {g}_{m, y}^{\alpha}) =  \left(\frac{2\pi \omega_0}{\varepsilon_{\infty}}\frac{\mu_{cv}^2}{\hbar}\right)^{1/2}\mathbf{Q}_m^{\alpha} \, ,
\label{coupling}
\end{equation}
which should be interpreted as the coupling strengths between the $m$-th mode of the PHC and the $\alpha$-th QD. To this end, we notice that the $2N$ equations in (\ref{alphbet}) can be solved only for those values of $\omega$ for which the $N\times N$ matrix 

\begin{widetext}
\begin{equation}
\Lambda_1 = 
\begin{pmatrix}
\omega^{1}_x - \omega - \sum_{m=1}^M \frac{g_{m,x}^{1} g_{m,x}^{1*}}{\omega_m - \omega} & - \sum_{m=1}^M \frac{g_{m,x}^{1} g_{m,y}^{1*}}{\omega_m - \omega} & \cdots & - \sum_{m=1}^M \frac{g_{m,x}^{1} g_{m,x}^{N*}}{\omega_m - \omega} & - \sum_{m=1}^M \frac{g_{m,x}^{1} g_{m,y}^{N*}}{\omega_m - \omega} \\
 - \sum_{m=1}^M \frac{g_{m,y}^{1} g_{m,x}^{1*}}{\omega_m - \omega} & \omega^{1}_y - \omega - \sum_{m=1}^M \frac{g_{m,y}^{1} g_{m,y}^{1*}}{\omega_m - \omega} & \cdots & - \sum_{m=1}^M \frac{g_{m,y}^{1} g_{m,x}^{N*}}{\omega_m - \omega} & - \sum_{m=1}^M \frac{g_{m,y}^{1} g_{m,y}^{N*}}{\omega_m - \omega} \\
\vdots & \cdots & \ddots & \cdots & \vdots \\
- \sum_{m=1}^M \frac{g_{m,y}^{N} g_{m,x}^{1*}}{\omega_m - \omega} & - \sum_{m=1}^M \frac{g_{m,y}^{N} g_{m,y}^{1*}}{\omega_m - \omega} & \cdots & - \sum_{m=1}^M \frac{g_{m,y}^{N} g_{m,x}^{N*}}{\omega_m - \omega}& \omega^{N}_y - \omega  - \sum_{m=1}^M \frac{g_{m,y}^{N} g_{m,y}^{N*}}{\omega_m - \omega}
\end{pmatrix}
\label{matrixone}
\end{equation}
\end{widetext}
is singular. This is a nonlinear equation, but we notice that it can be transformed into a more familiar form, since it is mathematically equivalent to finding the eigenvalues of the matrix 

\begin{equation}
\Lambda_2 = 
\begin{pmatrix}
\omega_{x}^1 & 0 & \cdots & 0 & g_{1,x}^1 & \cdots & g_{M, x}^1 \\
0 & \omega_{y}^1 & \cdots & 0 & g_{1,y}^1 & \cdots & g_{M, y}^1 \\
\vdots & \cdots & \ddots & \vdots & \vdots & \cdots & \vdots \\
0 & 0 & \cdots & \omega_{y}^N & g_{1,y}^N & \cdots & g_{M,y}^N \\
g_{1, x}^{1*} & g_{1, y}^{1*} & \cdots & g_{1, y}^{N*} & \omega_1 & \cdots & 0 \\
\vdots & \cdots & \ddots & \vdots & \vdots & \cdots & \vdots \\
g_{M, x}^{1*} & g_{M, y}^{1*} & \cdots & g_{M, y}^{N*} &0& \cdots & \omega_M  
\end{pmatrix}\, .
\label{fullmat}
\end{equation}
More precisely, solving $\det(\Lambda_1) = 0$ is equivalent to solving $\det(\Lambda_2 - \omega I_{(2N\times M)\times(2N\times M)}) = 0$, whenever $\omega \neq \omega_m \, \forall \, m = 1 \dots M$. The proof can be easily obtained by, on one hand, multiplying the equation for $\Lambda_1$ by $\prod_{m=1}^M (\omega_m - \omega)$, and on the other, using in the eigenvalue problem for $\Lambda_2$ the following identity for the determinant of a block-matrix:

\begin{equation}
\det\begin{pmatrix}A& B\\ C& D\end{pmatrix} = \det(D) \det(A - B D^{-1} C)\, .
\end{equation}

The poles $\omega = \omega_m$ will generally exist as solutions only when a photonic mode $\mathbf{Q}_m$ is fully decoupled from the system, i.e. when $\mathbf{g}_m^{\alpha} = 0 \, \forall \, \alpha$, in which case this mode can safely be excluded from the very beginning. The $2N + M$ complex eigenvalues of $\Lambda_2$ then define the frequencies (real part) and the loss rates ($-2 \times $ imaginary part) of the {\em polariton modes} of the system, while the eigenvectors 

\begin{equation}
\bm{\lambda} = (\lambda_{x}^1, \lambda_{y}^1, \dots \lambda_{x}^N, \lambda_{y}^N, \lambda_1 \dots, \lambda_M)
\label{eigvec}
\end{equation}
define the corresponding Hopfield coefficients\cite{Hopfield_1958}, which, for each eigenstate, give the probability amplitude of finding an excitation in the corresponding bare-exciton or bare-photon mode. Notice in addition that the matrix of Eq. (\ref{fullmat}) corresponds to a Tavis-Cummings Hamiltonian\cite{Tavis_1968} in the weak excitation regime, when only transitions from the ground state to the manifold of states with a single excitation are considered. Thus, notice that our approach has a straightforward extension to treating non-linear quantum dot dynamics, as the coupling constants in the off-diagonal terms of (\ref{fullmat}) can be used to write the Tavis-Cummings Hamiltonian in its most general from, i.e. including transitions among all excitation-number manifolds. This describes the system whenever the quantum dots behave as two-level systems, which is indeed the case for small dots under resonant excitation. 

The present formalism applies to a large variety of photonic structures and to an arbitrary spatial distribution of QDs. In this sense, it generalizes the results that were obtained for specific configurations\cite{Andreani_1999, Hughes_2004, Rao_2007, Lecamp_2007, Hughes_2007, Yao_2009, Kristensen_2011}. As an illustrating application, in section \ref{applications} we present results obtained for the case of two quantum dots embedded in several of the most widely studied photonic crystal structures: the $L3$ and $Ln$ cavities\cite{Akahane_2003}, and the $W1$ waveguide.

\section{Model parameters}

\label{parameters}

In order to quantify the susceptibility (\ref{susc}), we need an appropriate model of the exciton wave-function evaluated at equal electron and hole positions, $\Psi_{\alpha}(\mathbf{r}) = \Psi_{\alpha}(\mathbf{r}_e = \mathbf{r}, \mathbf{r}_h = \mathbf{r})$. This function is \textit{not} properly normalized as a function of $\mathbf{r}$ (the correct normalization is over $\mathbb{R}^3 \times \mathbb{R}^3$ as given in Eq. (\ref{psinorm})). In fact, similarly to the quantum well case \cite{Tassone_1990, Andreani_1994}, the oscillator strength of the exciton transition in the QD depends on the dimensionless quantity

\begin{equation}
C^2 = \left| \int \mathrm{d} \mathbf{r} \Psi_{\alpha}(\mathbf{r}) \right| ^2 \, .
\end{equation}

The particular \textit{shape} of the wave-function enters through the overlap integrals with the electric field, as given in Eq. (\ref{overlap}). As long as the size of the QDs is much smaller than the characteristic wavelength, the electric field varies very weakly in the region where $\Psi_{\alpha}$ is non-negligible, and thus the point dipole assumption, $\Psi_{\alpha} (\mathbf{r}) = C \delta(\mathbf{r} - \mathbf{r}_{\alpha})$, is a very good approximation. In what follows we will mostly use parameters typical of self-organized InGaAs QDs \cite{Bimberg_1999}, whose size lies in the $10-20 \mathrm{nm}$ range, with a typical exciton recombination energy of $1.3 \mathrm{eV}$ ($\lambda \approx 950 \mathrm{nm}$). For these values, we checked that assuming a Gaussian shape for $\Psi_{\alpha}(\mathbf{r})$ introduces little change with respect to the Dirac-delta assumption. Notice, however, that the strong dependence\cite{Gil_2002, Langbein_2004} of the QD oscillator strength with its size is still present, carried by the normalization constant $C$. One way to estimate this constant is through a microscopic model of $\Psi_{\alpha}(\mathbf{r})$\cite{Wang_1999, Stier_1999}. Here, instead, we take a more pragmatic approach, and compute $C$ based on the measured radiative decay rate of QDs. Following Ref. [\onlinecite{Parascandolo_2005}], this is given by twice the imaginary part of the quantity 

\begin{equation}
G_{\alpha} = i\frac{2\pi^2\mu_{cv}^2}{\hbar \varepsilon_{\infty}} \int_0^{\infty} \mathrm{d}k \vert \Psi_{\alpha\mathbf{k}} \vert^2 \frac{k(2k_0^2 - k^2)}{k_z} \, ,
\end{equation}
where $\Psi_{\alpha\mathbf{k}}$ is the Fourier transform of $\Psi_{\alpha} (\mathbf{r})$. With the assumption $\Psi_{\alpha} (\mathbf{r}) = C \delta(\mathbf{r} - \mathbf{r}_{\alpha})$, the decay rate is thus

\begin{equation}
\Gamma^{\alpha} = \frac{4}{3} \frac{ k_0^3}{\hbar \varepsilon_{\infty}} d^2 \, , 
\label{loss}
\end{equation}
where $k_0 = (\omega_0/c)\sqrt{\varepsilon_{\infty}}$, and we defined the dipole moment $d$ of the dot (also labeled  $\mathcal{D}$\cite{Thranhardt_2002} or $\mu$\cite{Khitrova_2006})  as 
\begin{equation}
d^2 = \mu_{cv}^2 C^2 \, .
\label{dipole}
\end{equation}
Eq. (\ref{loss}) coincides with the expression that is commonly adopted\cite{Andreani_1999, Thranhardt_2002, Khitrova_2006}. For typical QDs\cite{Hennessy_2007, Faraon_2008, Reinhard_2012}, with radiative lifetime of $1 \mathrm{ns}$ and exciton transition energy $\hbar \omega^{\alpha} \approx 1.3 \mathrm{eV}$, we obtain a squared dipole moment $d^2 \approx 0.51 \mathrm{eV \times nm^3}$.

The last requirement of the problem is the knowledge of the modes of the PHC structure, i.e. the set of orthonormal functions $\{\mathbf{Q}_m(\mathbf{r})\}$ and their corresponding eigenfrequencies $\omega_m$. Here, PHC modes are computed using the Bloch-mode expansion method \cite{Savona_2011}, which consists in expanding the modes on the basis of the Bloch modes of a regular waveguide. These latter were in turn computed using an expansion over the guided modes of a uniform dielectric slab.\cite{Andreani_2006} This approach turns out to be particularly well suited for elongated PHC cavities as considered in the present work. The computation was carried out over a finite supercell $S$ in the plane of the crystal, and infinite space along the orthogonal, $z-$direction. The orthogonality relation is then given by

\begin{equation}
\int_S \mathrm{d}^2\rho \int_{-\infty}^{\infty} \mathrm{d}z \; \mathbf{Q}_m(\bm{\rho}, z) \mathbf{Q}_n^*(\bm{\rho}, z) = \delta_{mn} \, .
\end{equation}

All the photonic crystals we consider are based on a triangular lattice of circular holes etched in a dielectric slab suspended in air. The specific parameters we chose are relevant to GaAs structures,\cite{Hennessy_2007, Faraon_2008,Reinhard_2012} namely: lattice constant $a = 260 \mathrm{nm}$, hole radius $65 \mathrm{nm}$, and slab thickness $120 \mathrm{nm}$, with a real part of the refractive index $\sqrt{\varepsilon_{\infty}}=3.41$. In this work we consider only ideal PHC structures, in the absence of any disorder that would arise from the fabrication process. Disorder in PHCs has two important effects. First, it determines extrinsic radiation loss rates of otherwise fully guided modes in waveguides, and strongly suppresses the quality factors of high-quality PHC cavities. This effect is here taken into account through the inclusion of a constant phenomenological loss rate for the modes under study, related to their quality factor $\mathcal{Q}$ by $\gamma = \omega/\mathcal{Q}$. For the $L3$ cavity of section \ref{L3}, we set $\mathcal{Q} = 10000$ or $30000$. For the longer $Ln$ cavities (section \ref{Ln}) and the $\mathit{W1}$ waveguide (section \ref{W1}), we set $\mathcal{Q} = 50000$ for all modes. The second way disorder affects the results is by modifying the spatial profiles of the electric field modes, especially in the case of waveguides. This effect lies beyond the scope of the  present work -- although a brief discussion is given in section \ref{discussion} -- and will be the object of a future work.

\section{Applications}

\label{applications}

In this section, we apply the formalism to the prototypical cases of one or two QDs embedded in elongated $Ln$ cavities or in a $\mathit{W1}$ waveguide.

\subsection{Application to an $\mathit{L3}$ cavity}
\label{L3}

The system of one quantum dot coupled to an $\mathit{L3}$ cavity has been widely studied\cite{Yoshie_2004, Hennessy_2007, Reinhard_2012} and is thus a good starting point for testing the present formalism. 

\begin{figure}[h]
\centerline{\includegraphics[width=8cm, trim = 1in 0in 0in 0in, clip = true]{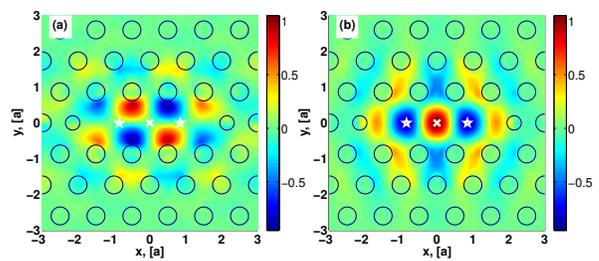}}
\caption{(Color online) Imaginary part of the electric field of the fundamental mode of an $\mathit{L3}$ cavity, (a): $\Im(Q_x(\mathbf{r}))$ and (b): $\Im(Q_y(\mathbf{r}))$. In the one-QD simulation, the dot was placed in the central maximum of the y-field (dot position marked by a white cross). For the two-QD simulations, the dots were placed in the corresponding secondary maxima (positions marked by white stars).}
\label{L3_profiles}
\end{figure}

The cavity is a modified $L3$ cavity\cite{Hennessy_2007, Reinhard_2012}, where the two holes on each side of the cavity are shifted outwards by $0.15a$, and their radii are decreased by $80\%$. This design improves the quality factor by more than one decade compared to that of a standard $L3$ cavity, while changing the field profile only marginally. We include in the computation only the fundamental cavity mode, shown in Fig. \ref{L3_profiles}. We further assume the QD to lie on the in-plane symmetry axis of the cavity, where $Q_x = 0$. The diagonalization of the matrix (\ref{fullmat}) is then equivalent to the well-known expression

\begin{equation}
\det \begin{pmatrix}
\omega_y - \omega & g_c \\
g^*_c & \omega_c  - \omega 
\end{pmatrix} = 0 \, .
\label{detL3}
\end{equation}

The coupling constant $g_c$, through Eq. (\ref{coupling}), is

\begin{equation}
g_c = \left(\frac{2\pi \omega_0}{\varepsilon_{\infty} \hbar}\right)^{1/2} d \,Q_y(\mathbf{r}_{\alpha}) \, ,
\end{equation}
which matches previous theoretical results \cite{Andreani_1999} when the dot is sitting in the center $\mathbf{r}_0$ of the cavity and the mode volume is defined as $\frac{1}{V} = \vert {Q}_y(\mathbf{r}_0) \vert^2$. As expected from Eq. (\ref{detL3}), for $ \vert g_c \vert^2 > \vert \gamma_c - \gamma_y\vert ^2 /16$, vacuum-field Rabi splitting appears between two polariton modes. The energy splitting at zero dot-cavity detuning is given by $2\hbar\Omega$, where the Rabi frequency $\Omega$ is

\begin{equation}
\Omega = \sqrt{\vert g_c \vert ^2 - \frac{(\gamma_c - \gamma_y)^2}{16}} \, .
\label{Rabi}
\end{equation}
Using the PHC and QD parameters we already introduced, the coupling constant was computed to be $\hbar \vert g_c \vert = 147 \mathrm{\mu eV}$, which compares perfectly with the most recently reported result for that system\cite{Reinhard_2012}.

After showing the way the standard one-dot cavity-QED results are reproduced with our formalism, we now proceed to the situation of two dots coupled to the same cavity mode (see Fig. \ref{L3_profiles}), and so radiatively coupled to each other. 

We assume a symmetric spatial configuration of the two dots with respect to the cavity center (see Fig. \ref{L3_profiles}), resulting in equal coupling constants $\hbar \vert g_c \vert = 125 \mathrm{\mu eV}$ for the two dots. Since usually $\gamma_c \gg \gamma_y$, i.e. the losses through the cavity mode are significantly larger than the QD losses through other channels (both non-radiative and radiative through modes other than the cavity mode), we set here and in all following sections $\gamma_y^{1,2} = 0$. Given the phenomenological way these rates enter the formalism, calculations can easily be generalized to include finite QD loss rates. Let us first consider the case of zero dot-dot detuning $\delta=\omega_y^1-\omega_y^2$. The relevant exciton states are in this case the symmetric and antisymmetric linear combinations of the two QD states, whose coupling to the cavity mode depends on the symmetry of the electric field profile. As discussed extensively in Ref. [\onlinecite{Portalupi_2011}], the $\mathit{L3}$ cavity symmetry is described by the $D_{2h}$ point group, and its fundamental mode belongs to the $B_{2u}$ irreducible representation, which is even with respect to the $\hat{\sigma}_{yz}$ symmetry operation (mirror reflection with respect to the $\mathit{yz}$ plane) -- as can also be seen from Fig. \ref{L3_profiles}. Hence, the antisymmetric QD state remains dark, while the symmetric one behaves as a single exciton with a coupling constant $\sqrt{2} g_c $. 

\begin{figure}[h]
\centerline{\includegraphics[width=8cm, trim = 0.7in 0in 0in 0.7in, clip = true]{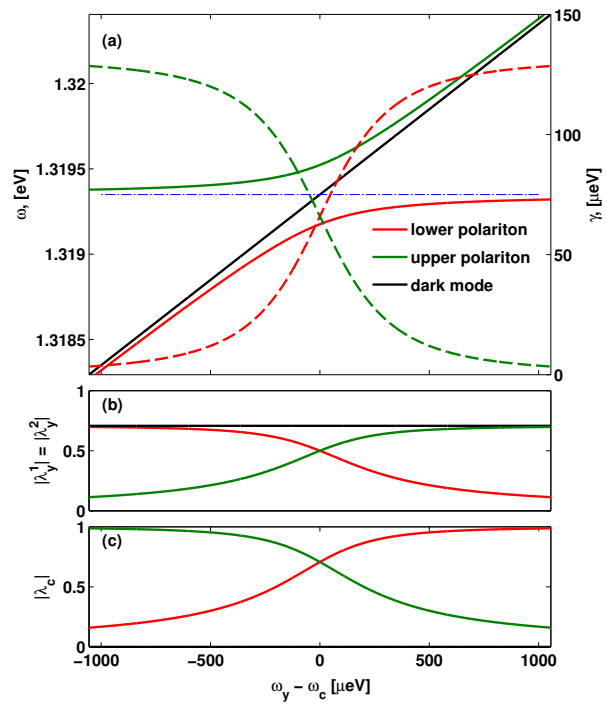}}
\caption{(Color online) (a): Eigenfrequencies (solid lines) and radiative rates (dashed lines) for two QDs with no dot-dot detuning, strongly coupled to an $\mathit{L3}$ cavity mode with $\mathcal{Q} = 10000$. With a dashed-dotted line, the bare cavity resonance is also indicated. The Hopfield coefficients for each solution, correspondingly color-coded, are presented in panels (b): equal (in absolute value) QD coefficients and (c): cavity coefficient.}
\label{L3_2D_det0}
\end{figure}

In Fig. \ref{L3_2D_det0} (a) we plot the eigenfrequencies of the system as a function of the detuning between the exciton resonance frequency $\omega_y$ (same for both dots) and the cavity resonance frequency $\omega_c$, as computed for a cavity quality factor of $\mathcal{Q} = 10000$. We observe vacuum Rabi splitting between an upper and a lower polariton in exactly the same way we would for a single dot coupled to the cavity, but in addition we see a dark mode which is a trivial solution, $\omega = \omega_y$. The splitting between the lower and the upper polaritons at zero dot-cavity detuning is $2 \hbar \Omega_c = 347 \mathrm{\mu eV}$, which for $\mathcal{Q} = 10000$ corresponds exactly to an effective coupling constant of  $\sqrt{2} \times 125\mathrm{\mu eV}$. The system is further characterized in panels (b) and (c), where we plot the Hopfield coefficients for each of the three eigenmodes (correspondingly color-coded). This clear collective behavior has been observed experimentally in a QD-cavity system\cite{Reitzenstein_2006, Laucht_2010, Kim_2011, Albert_2012}, while the more general dependence of the effective coupling constant with the number of coupled two-level systems $N$ -- given by $\sqrt{N} \vert g_c \vert$ -- has also been observed in a circuit-QED system\cite{Fink_2009}. It is very important to remark that this dependence has nothing to do with the $\sqrt{N} \vert g_c \vert$ energy splitting  of different rungs in a Jaynes-Cummings model, where $N$ would be the number of photons in the system: on the contrary, as discussed before, here we restrict to the linear response only, which holds in the limit of vacuum electromagnetic field. The effect in our case is simply due to the collective behavior of the $N$ resonant quantum dots.

\begin{figure}[h]
\centerline{\includegraphics[width=8cm, trim = 0.6in 0in 0in 0.5in, clip = true]{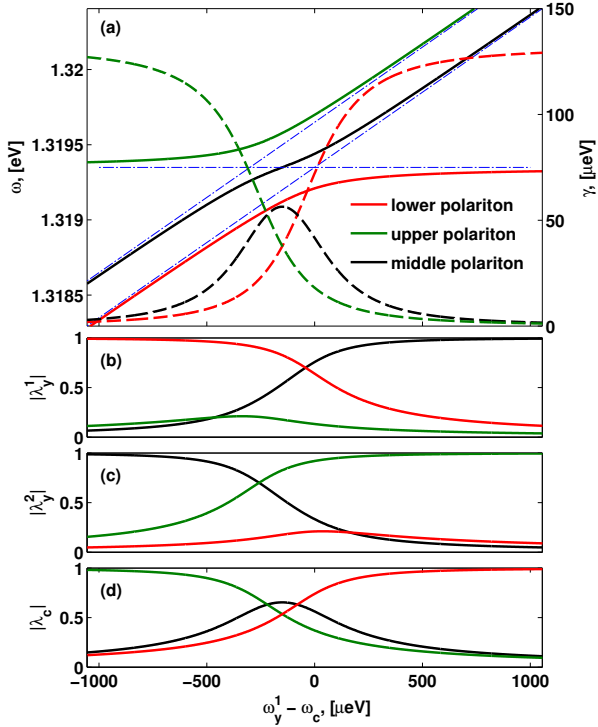}}
\caption{(Color online) (a): Eigenfrequencies (solid lines) and radiative rates (dashed lines) for two QDs with a dot-dot detuning of $300 \mathrm{\mu eV} $, strongly coupled to an $\mathit{L3}$ cavity mode with $\mathcal{Q} = 10000$. With dashed-dotted lines, the bare excitons and the bare cavity resonances are also shown. The Hopfield coefficients for each solution, correspondingly color-coded, are presented in panels (b): first exciton coefficient, (c): second exciton coefficient, and (d): cavity coefficient.}
\label{L3_2D_det300}
\end{figure}

The major experimental challenges to the radiative coupling of two spatially separated quantum dots is achieving both spatial control (to ensure strong overlap between each of the dots and the cavity mode) and spectral control (to ensure as small dot-dot and dot-cavity detuning as possible). Typically, QDs are characterized by an inhomogeneous distribution of exciton energies with a width of several meV. Then, two QDs are very likely to be detuned. In Fig. \ref{L3_2D_det300}, we study the same system, but assuming a detuning $\delta = 300 \mathrm{\mu eV}$. Close to resonance, all of the eigenmodes acquire a finite component from the cavity mode. Additionally, they have both a significant $\vert \lambda_y^1 \vert$ coefficient (panel (b)), and a significant $\vert \lambda_y^2 \vert$ coefficient (panel (c)), implying that there is a sizable radiative coupling present. The radiative coupling is expected to vanish as the cavity-dot detunings become much larger than the coupling constant, and an expression for an effective coupling strength in this limit was derived in Ref. [\onlinecite{Imamoglu_1999}] and [\onlinecite{Gywat_2006}]. Concerning the spatial control, it is important to note that our approach allows for a statistical analysis of the effect of an imperfect positioning of the dots, although such an analysis lies beyond the scope of the present work. 

\begin{figure}[h]
\centerline{\includegraphics[width=8cm, trim = 0.7in 0in 0in 0.5in, clip = true]{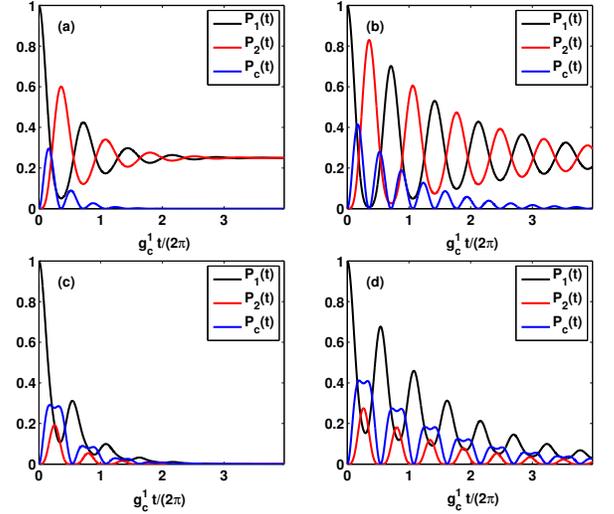}}
\caption{(color online) Time evolution of the probability of an excitation in one dot to be transferred to the second dot or to the cavity. (a): $\delta = 0$, $\mathcal{Q} = 10000$; (b): $\delta = 0$, $\mathcal{Q} = 30000$; (c) $\hbar \delta = 300\mathrm{\mu eV}$, $\mathcal{Q} = 10000$; (d) $\hbar \delta = 300\mathrm{\mu eV}$, $\mathcal{Q} = 30000$.}
\label{time_L3}
\end{figure}

We now address the question of how the excitation transfer process depends on time. This aspect is of particular importance to assess the usefulness of the radiative excitation transfer as a coupling mechanism between different qubits in a semiconductor-based quantum gate architecture. In the present case, when polaritonic features are spectrally resolved, one correspondingly expects the excitation to oscillate between the different basis states, including the photon state. To illustrate this aspect, we compute the time-dependent amplitudes of the various basis states, assuming that one QD is excited at $t=0$. From these amplitudes, we extract time-dependent probabilities of finding the excitation in each of the basis modes, expressed in vector form as

\begin{equation}
\mathbf{P}(t) = \left|e^{-i\Lambda_2 t} \bm{\lambda}_{in} \right|^2 \, ,
\label{prob}
\end{equation}
where $\Lambda_2$ is the matrix of Eq. (\ref{fullmat}). These probabilities are properly normalized if one accounts also for the probability $P_{out}(t)$ of the excitation to have radiated out of the system, i.e. $\sum P_i(t) = 1 - P_{out}(t)$. In Fig. \ref{time_L3} we plot these time-resolved probabilities for a starting excitation in one of the QDs, i.e. $\bm{\lambda}_{in} = (1, 0, 0)$. We study four different cases: either zero dot-dot and dot-cavity detuning, or $\hbar \delta = 300 \mathrm{\mu eV}$ (with the cavity frequency tuned at the average of the two exciton frequencies), and cavity $\mathcal{Q}$-factor equal to either $10000$ or $30000$. In panels (a) and (b), where $\delta = 0$, the probabilities never decay to zero due to the presence of a dark state and the fact that no non-radiative decay mechanism was included. In panels (c) and (d) a dark state no longer exists, and a clear decay of the excitation with a characteristic lifetime depending on the $\mathcal{Q}$-factor is visible. All plots show that the excitation oscillates between the three possible states, on a time scale defined through the radiative coupling strength. In particular, the probability of finding the system in an excited state of the second QD remains sizable over several oscillation periods, showing that a significant dot-dot interaction can be achieved with experimentally feasible parameters. 
These results generally agree with specific setups of radiatively coupled QDs in photonic crystals, that have been recently studied in the literature.\cite{Yao_2009, Kristensen_2011}

\subsection{Application to $Ln$ cavities}
\label{Ln}

Recently, using $Ln$ cavities with $n > 3$ to achieve light-matter coupling has spurred interest\cite{Choi_2007, Felici_2010, Surrente_2011}, as these cavities generally have a larger quality factor than the $\mathit{L3}$ ones -- though at the expense of a larger mode volume and thus a smaller dot-cavity coupling strength. 

\begin{figure}[h]
\centerline{\includegraphics[width=8cm, trim = 1.8in 0in 0in 0in, clip = true]{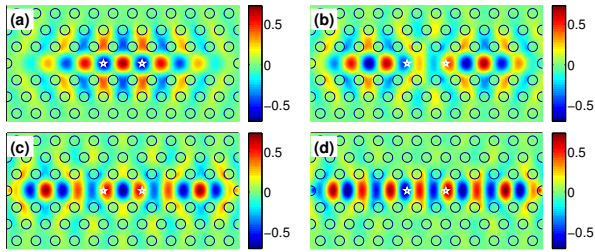}}
\caption{(Color online) $\Im(Q_y(\mathbf{r}))$ for the four lowest-energy modes of the $\mathit{L11}$ cavity; (a): Fundamental mode $M_1$, at $\hbar \omega_1 = 1.3065 \mathrm{eV}$, (b): $M_2$, $\hbar \omega_2 = 1.3125 \mathrm{eV}$, (c): $M_3$, $\hbar \omega_3 = 1.3269 \mathrm{eV}$, (d): $M_4$, $\hbar \omega_4 = 1.3565 \mathrm{eV}$. The positions of the quantum dots are marked with white stars.}
\label{L11_profiles}
\end{figure}

Here, we investigate cavities of varying length $\mathit{n}$ with a common setup, illustrated in Fig. \ref{L11_profiles} for $\mathit{n} = 11$. In the figure, we show the first four modes, $M_{1-4}$, of the $\mathit{L11}$ cavity, with resonant energies $1.3065 \mathrm{eV}, 1.3125    \mathrm{eV}, 1.3269 \mathrm{eV}, $ and $1.3565\mathrm{eV}$, respectively. In all the results to follow, for all $\mathit{n}$, the two dots were placed in the center of an elementary cell on each side of the center of the defect (i.e. at a distance $a$ from the center of the cavity and so $2a$ from each other), where the coupling constants for each of them in the $\mathit{n} = 11$ case are $\vert \hbar g_1 \vert = 94 \mathrm{\mu eV}$, $\vert \hbar g_2 \vert = 55 \mathrm{\mu eV}$, $\vert \hbar g_3 \vert = 65 \mathrm{\mu eV}$, and $\vert \hbar g_4 \vert = 89 \mathrm{\mu eV}$. Since the smallest energy difference between the cavity resonances in this case is between $\omega_1$ and $\omega_2$, and is $\approx 6 \mathrm{meV}$, i.e. much larger than all the coupling strengths, it is reasonable to expect that the dots will never couple significantly to more than one mode. Thus, the phenomenology of the system will be, qualitatively, the same as the one described in section \ref{L3}, which was also verified by our computations.

The situation should change significantly when increasing the length $n$ of the photonic defect. Then, we expect the energy spacing between the resonant frequencies of the $Ln$ cavity to decrease and eventually become comparable to the typical coupling strength. In this situation, the radiative transfer process is no longer mediated by an isolated cavity mode, and a smooth transition to a multi-mode coupling regime is expected. In order to determine at which cavity length this crossover occurs, one should also consider the fact that the coupling of a dot to each individual mode decreases with the increase of the mode volume. As a result, the crossover length is increased with respect to what would be given by a simple assumption of constant coupling strength per mode.

\begin{figure}[h]
\centerline{\includegraphics[width=8cm, trim = 1in 0in 0in 0.8in, clip = true]{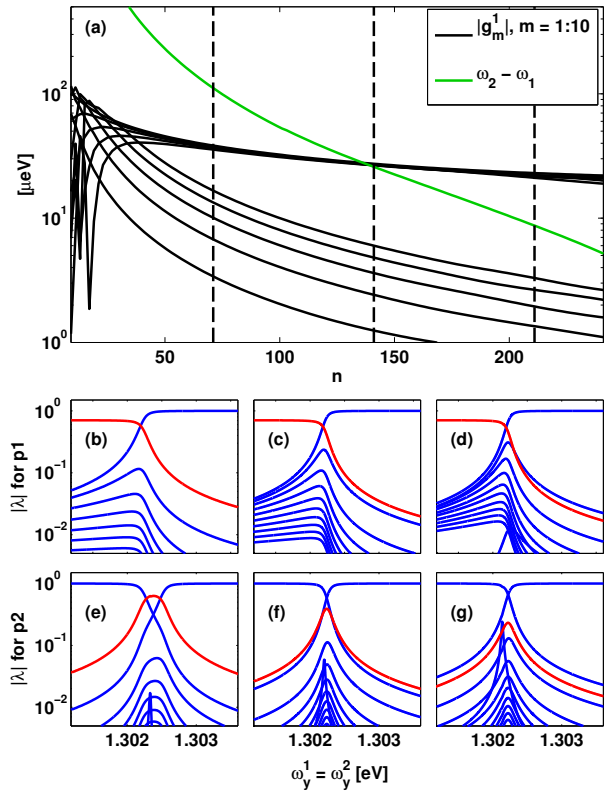}}
\caption{(Color online) (a): Black lines -- coupling constants between one QD and the ten lowest modes of an $Ln$ cavity, vs. $n$; green line -- energy separation between the lowest two cavity modes. (b)-(d): Hopfield coefficients of one polariton eigenstate as a function of the bare exciton frequency $\omega^1_y$ with no dot-dot  detuning, for $n = 71$, $n = 141$, $n = 211$ (the values marked by dashed vertical lines in (a)). The red line shows the dot coefficients, while the blue lines belong to the many cavity modes. (e)-(g): Same as (b)-(d) but for another polariton state.}
\label{Ln_comparison}
\end{figure}

In Fig. \ref{Ln_comparison}, we plot the minimum mode-separation $\omega_2 - \omega_1$ vs. the length $n$ of the cavity, and in addition show the coupling strengths $\vert g_m^{1} \vert$ for $m = 1\dots 10$. For all $n$, the dots were placed as in Fig. \ref{L11_profiles} -- at a distance $a$ on each side of the center of the cavity. The fact that half of the coupling constants decay much faster as a function of $n$, is again explained by the particular symmetry of the field profiles. It turns out that for every $n$, the modes alternate between symmetric and antisymmetric w.r.t. $\hat{\sigma}_{yz}$, as can be seen in Fig. \ref{L11_profiles} for the $\mathit{L11}$ case. In the limit of large $n$, the antisymmetric modes have a small amplitude at the QD positions close to the node, resulting in a small radiative coupling strength. 

The crossover from a single-mode to a many-mode regime occurs around $n = 150$, as clearly visible in Fig. \ref{Ln_comparison}. In panels (b)-(g), we show the corresponding Hopfield coefficients for three different values of $n$, given by $n=71$, $n=141$, and $n=211$, also indicated by dashed lines in panel (a), and for two different polariton modes.
Consequently, for $n = 71$, the Hopfield coefficients of two different polariton eigenstates, shown in panels (b) and (e) respectively, are still largely dominated by one cavity and one dot component. On the other hand, for $n = 211$ -- panels (d) and (g) -- the value of several photonic fractions $\lambda_m$ is non-negligible. 

\begin{figure}[h]
\centerline{\includegraphics[width=8cm, trim = 0.8in 0in 0in 0.8in, clip = true]{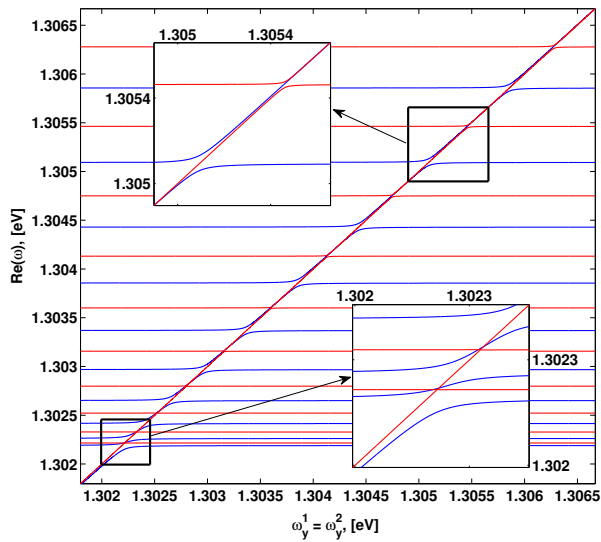}}
\caption{(Color online) Eigenfrequencies for two QDs with no dot-dot detuning in an $\mathit{L141}$ cavity ($\mathcal{Q} = 50000$ for each mode) vs. the resonant frequency of the excitons. The insets show close-ups over two selected regions.}
\label{L141_2QD_det0}
\end{figure}

In Fig. \ref{L141_2QD_det0} we plot the polariton energies as a function of QD-exciton energy in the case of the $\mathit{L141}$ cavity, for $\delta = 0$. As mentioned already, the photon modes alternate between symmetric and anti-symmetric w.r.t. the $\hat{\sigma}_{yz}$ operator, hence coupling to either the symmetric or anti-symmetric linear combination of the QD states is present. In the figure, the polaritons due to a combination of symmetric states are denoted by blue lines, while the anti-symmetric combinations are represented by red lines. In the symmetric case, the exciton-photon coupling strength is always large, and anti-crossing occurs at every mode. In the anti-symmetric case, the results show a transition from weak coupling (close to the lowest $\omega_m$) to strong coupling (anti-crossing is visible in the higher-$\omega$ inset), due to the fact that the coupling strengths there become larger than $\omega_m/4\mathcal{Q}$. It is clear both from Figs. \ref{Ln_comparison} and \ref{L141_2QD_det0} that for $n\rightarrow\infty$, the dots couple to a  structured continuum of photon modes, reproducing the physics of a W1 waveguide. This situation is studied in the next section. 

\subsection{Application to a W1 waveguide} 
\label{W1}

The results obtained for a $Ln$ cavity indicate that radiative coupling is still sizable in very long structures and might be effective even at very long distance between the two QDs. Here, we investigate this possibility in more detail, by considering QDs embedded in a W1 photonic crystal waveguide. 

Coupling of a single dot to a W1 waveguide (or a similar structure) with the purpose of spontaneous emission enhancement (and the potential application as a single-photon source) has already been widely discussed theoretically\cite{Rao_2007, MangaRao_2007, Lecamp_2007}, and achieved experimentally\cite{Viasnoff-Schwoob_2005, Lund-Hansen_2008, Thyrrestrup_2010, Schwagmann_2011, Hoang_2012, Laucht_2012, Laucht_2012a}. The fact that it is already possible to couple efficiently a dot to the guided modes of the waveguide is promising in view of achieving radiative coupling between \textit{two} dots that could -- due to the spatial extension of the structures and the modes they support -- extend to inter-dot distances for which targeting each dot individually by a laser pulse is possible. 

\begin{figure}[h]
\centerline{\includegraphics[width=8cm, trim = 1in 0in 0in 1in, clip = true]{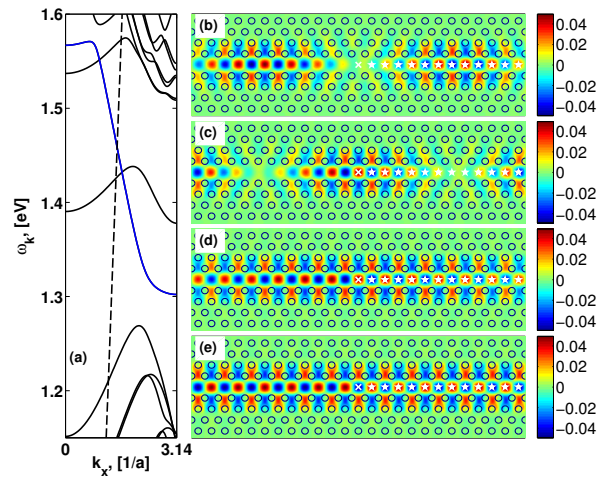}}
\caption{(Color online) (a): Band structure of the $W1$ waveguide; the dashed line shows the light cone. The QD resonant energies are close to the band-edge energy of the main guided band (blue). The field profiles of four guided modes in that spectral region are shown, over a small stretch of the waveguide: (b), (c): the two degenerate modes at $\hbar \omega_k = 1.30308\mathrm{eV}$ (anti-symmetric combination in (b), symmetric in (c)); (d): the symmetric mode at $\hbar \omega_k = 1.30224\mathrm{eV}$ (e): the symmetric mode at $\hbar \omega_k = 1.30218\mathrm{eV}$ (at the band edge). In all computations, one QD was placed in the center of the waveguide (white cross), while the second one was placed in the center of one of the successive elementary cells (white stars).}
\label{W1_profiles}
\end{figure}

We begin our study by looking at the modes of the W1 waveguide. The band structure is presented in Fig. \ref{W1_profiles} (a), where two guided bands in the band-gap of the regular crystal are visible. Strongest dot-PHC coupling is typically achieved for the smallest group velocity (largest local density of states of the photonic modes), and so the spectral range we concentrate on is around the band-edge of the main guided band (blue line), where the group velocity of the ideal photonic structure vanishes. The second guided band is spatially odd with respect to a $\hat{\sigma}_{xz}$ reflection,\cite{Andreani_2006} and would not couple to the exciton state of a QD located at the center of the waveguide. In the simulations below, we compute the $W1$ modes for $2048$ $k$-points in the interval $(-\pi/a, \pi/a]$, which is equivalent to simulating a waveguide of length 2048 elementary cells with periodic boundary conditions (PBC). In panels (b)-(e) of Fig. \ref{W1_profiles}, we show the electric field profiles of four modes lying close in energy to the band edge of the main guided band. As is the case with all structures we considered so far, this band has vanishing $Q_x$ component on the symmetry axis of the waveguide, allowing us again to include the $y$-polarized fields only. Furthermore, modes at $\pm k$ are degenerate -- one propagating and one counter-propagating -- with real-space profiles proportional to $\exp(ikx)$ and $\exp(-ikx)$, respectively. As basis states, we take the symmetric and the anti-symmetric combination of the degenerate guided modes, representing the fields by their ``standing wave'' profiles: one with a maximum and one with a zero amplitude in the center of the guide (compare panels (b)-(c)). Without loss of generality (due to the PBC), we place one dot at that position, so that it couples to one of the modes only. The second dot is then placed in the center of a successive elementary cell, and will, in general, couple to every mode in the basis thus constructed, so even for zero dot-dot detuning, no fully dark state is present.

\begin{figure}[t]
\centerline{\includegraphics[width=8cm, trim = 0.6in 0in 0in 0.6in, clip = true]{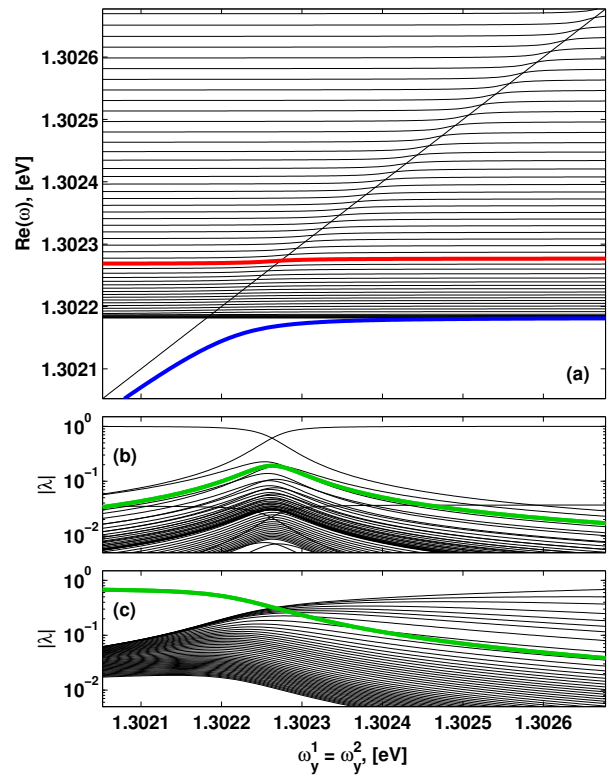}}
\caption{(Color online) (a): Eigenfrequencies for two QDs in the waveguide, with no dot-dot detuning. The Hopfield coefficients for the red line of (a) are shown in (b), where the green line shows the QD coefficients $\vert \lambda_y^1 \vert = \vert \lambda_y^2 \vert$, and the black lines show all the waveguide coefficients $\vert \lambda_m \vert$. The same in (c), but for the blue line of (a). The loss rate of each mode is $\gamma_{\mathrm{w}} = \omega_y^1/50000 \approx 26 \mathrm{\mu eV}$.}
\label{W1_2QD_det0}
\end{figure}

\begin{figure}[t]
\centerline{\includegraphics[width=8cm, trim = 0.6in 0in 0in 0.8in, clip = true]{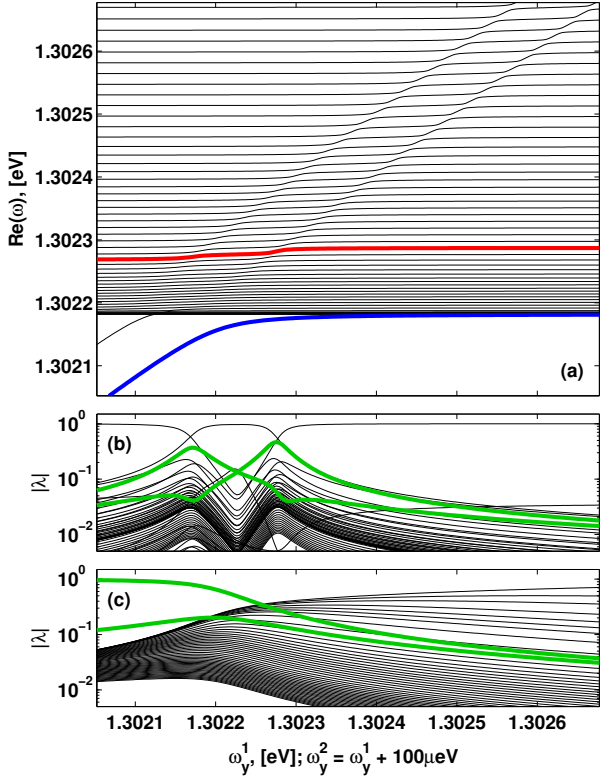}}
\caption{(Color online) (a): Eigenfrequencies for two QDs in the waveguide, with dot-dot detuning $\hbar \delta = 100 \mathrm{\mu eV}$ and $\gamma_{\mathrm{w}}$ as in Fig. \ref{W1_2QD_det0}. The Hopfield coefficients for the red line of (a) are shown in (b), where the two green lines show the QD coefficients $\vert \lambda_y^{1, 2} \vert$, and the black lines, all the waveguide coefficients $\vert \lambda_m \vert$. Same in (c), but for the blue line of (a).}
\label{W1_2QD_det100}
\end{figure}

In Fig. \ref{W1_2QD_det0} (a), we show the polariton structure in the spectral range close to the band edge of two dots with dot-dot detuning $\delta = 0$, with the second dot placed at the closest possible distance, $a$, from the first one. The quality factor of each of the photonic modes was again set to $\mathcal{Q} = 50000$ for all modes. While a strong dependence of the $W1$ loss rates on the group velocity close to the band edge has been shown in transmission measurements\cite{OFaolain_2007}, this dependence is heavily influenced by back-scattering due to disorder. In our case, we model stationary modes rather than transport, and the only relevant radiative loss is the one out of the plane of the PHC slab. Then, the assumption of approximately constant $\mathcal{Q}$-s is realistic, as seen from microscopic modeling of extrinsic disorder-induced losses \cite{Savona_2011, Gerace_2004}. Polariton modes originating from antisymmetric photon modes are essentially uncoupled and are not displayed (although, they were still included in the computation). The coupling constants of each of the dots to each of the symmetric modes varies very little, and is $\hbar \vert g_m^{1,2} \vert \approx 7 \mathrm{\mu eV}$. The $\omega = \omega_y^1$ solution (straight diagonal in panel (a)) is due to the anti-symmetric QD combination, which is almost dark. The strongest anti-crossing behavior is exhibited by the polariton lying below the band edge (blue line), whose Hopfield coefficients are given in panel (c). The remaining polariton modes display similar behavior, so the Hopfield coefficients of just one of them (the red line of (a)) are given in panel (b). For completeness, the same plots but for $\hbar \delta = 100 \mathrm{\mu eV}$ are given in Fig. \ref{W1_2QD_det100}. In this case, no dark modes are present and Hopfield coefficient corresponding to the two QDs are generally different from each other, as seen in panels (b) and (c). In both Figs.  \ref{W1_2QD_det0} and \ref{W1_2QD_det100} we observe that anti-crossings are still present -- though characterized by a very small energy splitting -- where the exciton becomes resonant with the various guided modes. This situation can be understood as the precursor to the structured continuous spectrum of modes that would arise in the limit of infinite waveguide length, analogous, for example, to the polariton modes arising from the interaction between an exciton in a two-dimensional quantum well and the three-dimensional continuum of electromagnetic modes.\cite{Tassone_1990, Tassone_1992}

The present formalism provides a detailed quantitative account of the effect of the guided electromagnetic field on the radiation properties of few QDs. In particular, we derive below the Purcell enhancement of the radiative rate characterizing a single QD, and the distance-dependence of the radiative excitation transfer process between two distant QDs. We compute these properties both numerically, and analytically. To this purpose, let us consider the elements of the matrix $\Lambda_1$ introduced by Eq. (\ref{matrixone}):

\begin{equation}
\Lambda_1^{\alpha \beta} = (\omega^{\alpha} - \omega)\delta_{\alpha \beta} - G^{\alpha \beta}(\omega) \, ,
\end{equation}
where the coupling matrix elements $G^{\alpha \beta}$ are proportional to the Green's function of eq. \ref{green}:

\begin{equation}
G^{\alpha \beta}(\omega) = \sum_{m=1}^M \frac{g_{m}^{\alpha} g_{m}^{\beta *}}{\omega_m - \omega} = d^2 \frac{2 \pi}{\epsilon_{\infty} \hbar} \frac{\omega^2}{c^2} G(\mathbf{r}_{\alpha}, \mathbf{r}_{\beta}, \omega).
\label{lamab}
\end{equation}
For a structure with no sharp resonances -- like the waveguide -- we can take advantage of the exciton-pole approximation and substitute $\omega = \omega_0$ in the denominator, in which case $G^{\alpha \alpha}(\omega_0)$ is the self-interaction energy of each dot, while $|G^{12}(\omega_0)|$ is an effective coupling constant for the case of two dots with zero dot-dot detuning, i.e. $\omega^1_y = \omega^2_y = \omega_0$. In order to derive an analytical expression for the coupling, let us replace the sum with an integral over $k = k_x$, and use the fact that, in accordance with Bloch's theorem, when $\mathbf{r}_{\alpha}$ and $\mathbf{r}_{\beta}$ are in the center of an elementary cell, $g_k(\mathbf{r}_{\beta}) = \exp{(-\mathrm{i}kx)} g_k(\mathbf{r}_{\alpha})$, to write

\begin{equation}
G^{\alpha \beta}(\omega_0) = \frac{a}{2\pi} \int_{-\frac{\pi}{a}}^{\frac{\pi}{a}}\mathrm{d}k \frac{\left| g_k \right|^2 e^{\mathrm{i}kx}}{\omega(k) - \omega_0} \, .
\end{equation}

A few simplifications are due. First, we write $\omega_k = \Re(\omega(k))$ and $\gamma_{\mathrm{w}} = -2\Im(\omega(k))$, and assume the latter is constant, equal to $\omega_0/\mathcal{Q}$. Furthermore, we assume $\left| g_k \right|^2 = \left| g \right|^2$, i.e. the coupling strength has weak dependence on $k$. This feature is due to the small spatial extension of the exciton wave function, resulting in a very broad distribution in Fourier space with approximately constant overlap with all guided modes, and is also confirmed by our numerical results. Finally, by taking $k_0$ as the positive Bloch momentum for which the guided mode is resonant with the exciton frequency $\omega_0$, and defining the corresponding group velocity

\begin{equation}
v_g = -\left. \frac{\mathrm{d} \omega_{k}}{\mathrm{d} k} \right|_{k_0} \, ,
\end{equation} 
we get

\begin{equation}
G^{\alpha \beta}(\omega_0) \approx \frac{a}{\pi} \frac{\left| g \right|^2}{v_g} \int_{0}^{\frac{\pi}{a}}\mathrm{d}k \frac{ \cos{(kx)}}{k - k_0 - \mathrm{i}\frac{\gamma_{\mathrm{w}}}{2v_g}} \, . 
\label{lamapprox}
\end{equation}

This expression holds in the limit where the resulting spectral linewidth is small enough so that the group velocity is still well defined. It can now be applied for example to obtain the radiative lifetime of a single dot embedded in the waveguide as $\Gamma^{\alpha} = 2\Im(G^{\alpha \alpha})$, and so

\begin{equation}
\Gamma^{\alpha} = \frac{2a}{\pi}\frac{\left| g \right|^2}{v_g} \left. \tan^{-1}{\left(\frac{2(k-k_0)v_g}{\gamma_{\mathrm{w}}}\right)}\right|_{k = 0}^{k = \pi/a} \, .
\label{w1loss}
\end{equation}
The Purcell factor for the enhancement of the single-dot spontaneous emission rate is then given by the ratio between Eq. (\ref{w1loss}) and Eq. (\ref{loss}). This result takes into account the detailed structure of the photonic environment resulting from the waveguide. In this respect, it generalizes the result obtained by assuming that only one Bloch mode at wave vector $k=k_0$ determines the radiation loss process.\cite{Rao_2007, Lecamp_2007} This simplified result is recovered by taking the limit $\gamma_{\mathrm{w}} \rightarrow 0$ in the integral (\ref{lamab}), namely by assuming that the guided Bloch mode has vanishing extrinsic radiation loss rate. The emission rate $\Gamma^{l}$ of the dot into leaky modes can also be estimated numerically by restricting the summation in Eq. (\ref{lamab}) to the modes which lie above the light-cone only. Then, the $\beta$-factor in the absence of non-radiative decay mechanisms can also be computed as

\begin{equation}
\beta = \frac{\Gamma^{\alpha}}{\Gamma^{\alpha} + \Gamma^{l}} \, ,
\end{equation}
and a further generalization to the case in which non-radiative processes are also present follows straightforwardly.

As a development from the previous works which consider just one dot in the waveguide, we now proceed to quantify the radiative excitation transfer process between \textit{two} QDs and its dependence on inter-dot distance. The closed-form expression for the cross-coupling term $G^{12}$, obtained by carrying out the integral (\ref{lamapprox}) reads

\begin{figure}[h]
\centerline{\includegraphics[width=8cm, trim = 1.5in 0in 0in 1in, clip = true]{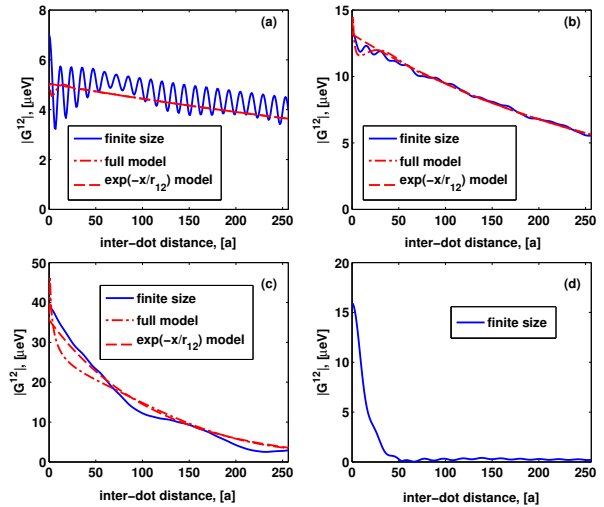}}
\caption{(Color online) Absolute value of the off-diagonal term of the matrix in Eq. (\ref{matrixone}), in the exciton-pole approximation, computed numerically for a finite-size waveguide (solid line), analytically through Eq. (\ref{lam12}) (dashed-dotted line), and through an exponential decay model with characteristic distance $r_{12} = 2v_g/\gamma_{\mathrm{w}}$ (dashed line, $\gamma_{\mathrm{w}}$ as in Fig. \ref{W1_2QD_det0}). In (a): $\hbar \omega_0 = 1.30353\mathrm{eV}$, $n_g = 74$, (b): $\hbar \omega_0 = 1.30240\mathrm{eV}$, $n_g = 195$, (c): $\hbar \omega_0 = 1.30220\mathrm{eV}$, $n_g = 525$, (d): $\hbar \omega_0 = 1.30208 \mathrm{eV}$, i.e. 100 $\mathrm{\mu eV}$ below the band edge.}
\label{W1_distance}
\end{figure}

\begin{widetext}
\begin{equation}
G^{12} = \frac{a}{\pi}\frac{\left| g \right|^2}{v_g} \left. \left[\cosh\left(\frac{x}{r_{12}} - \mathrm{i}k_0 x\right) \mathrm{Ci}\left(-\mathrm{i}\frac{x}{r_{12}} + (k - k_0)x\right) + \mathrm{i} \sinh\left(\frac{x}{r_{12}} - \mathrm{i}k_0 x\right) \mathrm{Si}\left(\mathrm{i}\frac{x}{r_{12}} - (k - k_0)x\right)\right]  \right|_{k = 0}^{k = \frac{\pi}{a}} \, ,
\label{lam12}
\end{equation} 
\end{widetext}
where $\mathrm{Ci}(z)$ and $\mathrm{Si}(z)$ are respectively the cosine integral and the sine integral functions, and we defined $r_{12} = 2v_g/\gamma_{\mathrm{w}}$. The quantity $r_{12}$ is simply the decay length associated to the propagation of light along the resonant guided mode. We expect this decay to characterize also the distance dependence of the radiation transfer process. Indeed, under the ideal assumption of vanishing radiation loss rate for the guided mode, in a one-dimensional geometry one would expect the radiative transfer process to be \textit{independent of the distance}. For comparison, as has already been shown, the coupling strength decays as $R_{\alpha \beta}^{-1}$ in 3D bulk semiconductor\cite{Parascandolo_2005}, and as $R_{\alpha \beta}^{-1/2}$ in a 2D planar cavity system\cite{Tarel_2008}. In Fig. \ref{W1_distance}, we display the absolute value of $G^{12}$ computed numerically through Eq. (\ref{lamab}), for four different values of the exciton frequency $\omega_0$ of the two QDs, in a waveguide of length $2048a$. This quantity is compared to the result obtained from the analytical model of Eq. (\ref{lam12}) and to the simpler assumption of an exponential dependence $|G^{12}| = |G^{11}| e^{-x/r_{12}}$. In panel (d), where $\omega_0$ is taken to lie below the edge of the guided band, the group velocity cannot be properly defined, and thus the analytical model does not apply.

Apart from this case, it is clear that the distance dependence of the inter-dot coupling is perfectly captured by the simple exponential decay model. The oscillations of the numerical curve in panel (a) are due to the finite length of the waveguide and reproduce the spatial behavior of the Bloch mode at $k=k_0$ that dominates the transfer process. These oscillations cannot obviously be reproduced by the analytical model that implicitly assumes an infinitely extended waveguide. As anticipated, the numerical results show that the distance dependence of the transfer rate is expressed by the decay associated to the light propagation, and quantified by the decay length $r_{12}$. It is interesting to note that even for very small group velocities, e.g. $v_g < c/500$, the interaction distance is still of the order of $100a = 26 \mu \mathrm{m}$, i.e. of mesoscopic scale, thus confirming the potential of the $\mathit{W1}$ for very long-distance dot-dot coupling. 

More generally, Eq. (\ref{lam12}) suggests that there is a compromise, enforced by the group velocity, between strength and distance dependence of the transfer process. The overall strength of the transfer rate depends inversely on the group velocity. This expresses the magnitude of the local density of states at the QD exciton frequency or, in a more suggestive picture, the fact that slow light interacts with a QD over a longer time lapse. However, a smaller group velocity also implies a shorter characteristic decay length $r_{12}$, as we are assuming a constant radiation loss rate. In a realistic system, including disorder, we further expect the group velocity picture to break down at frequencies close to the band edge, where disorder-induced localization of light dominates and the spatial decay associated to the localization length becomes shorter than $r_{12}$. This calls for an analysis including disorder effects, that we will consider in a future work.

\begin{figure}[h]
\centerline{\includegraphics[width=8cm, trim = 1.1in 0in 0in 0.8in, clip = true]{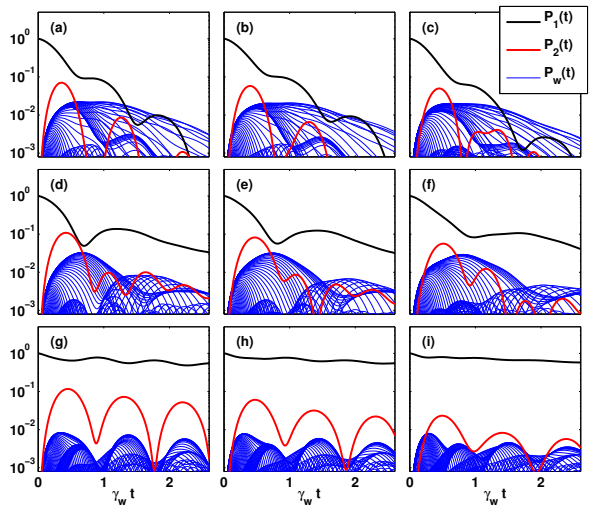}}
\caption{(Color online) Time evolution of the probability of an excitation in one dot to be transfered to the second dot or to the many PHC modes (blue lines). The dot-dot detuning is $\hbar \delta = \mathrm{100 \mu eV}$. Horizontally across the panels, the inter-dot distance changes from $260 \mathrm{nm}$ to $2.6 \mathrm{\mu m}$ to $5.2 \mathrm{\mu m}$. Vertically across the panels, the exciton frequency of the first dot changes from (a)-(c): $\hbar \omega_y^1 = 1.30224\mathrm{eV}$ (close to the band edge), through (d)-(f): $\hbar \omega_y^1 = 1.30218\mathrm{eV}$ (at the band edge), to (g)-(i): $\hbar \omega_y^1 = 1.30208\mathrm{eV}$ ($100 \mathrm{\mu eV}$ below the band edge energy, which is then resonant with $\hbar \omega_y^2$).}
\label{W1_timedep}
\end{figure}

We conclude this section by studying the time-dependent probability amplitudes of the excitation lying in each mode. These quantities are plotted in Fig. \ref{W1_timedep}, assuming that one QD is excited at $t=0$, for three different inter-dot distances and three different values of $\omega_y^1$. In all cases, $\hbar \delta = 100 \mathrm{\mu eV}$ was imposed. As discussed above, the transfer mechanism is driven by several light modes. The plots show that the radiative transfer process still occurs and, in particular, the marked oscillations are characterized on average by a period that can be associated to an effective transfer rate $\hbar \Omega = 50-60 \mathrm{\mu eV}$. As in short $Ln$ cavities, this rate is quite sizable and should be observable in state-of-the-art GaAs-based photonic structures.\cite{Viasnoff-Schwoob_2005, Surrente_2011, Lund-Hansen_2008, Thyrrestrup_2010, Schwagmann_2011, Hoang_2012, Laucht_2012, Laucht_2012a}

\section{Discussion and Outlook}
\label{discussion}

We have developed a general formalism of linear radiation-matter coupling in systems of many QDs embedded in a photonic crystal structure. The formalism is an extension of the exciton-polariton formalism well known for bulk semiconductors and quantum wells. It provides a quantitative account of a variety of radiative effects, starting from the basic microscopic parameters of the QD-PHC system. It is important to establish a relation between the present approach and  previous works that use the photonic Green's function\cite{Hughes_2004, Rao_2007, Hughes_2007, Yao_2009, Kristensen_2011}. The equations obtained there have the advantage of highlighting the importance of each single mode in determining the effects under study, but, on the other hand, incorporate either single-mode approximations or perturbative expansions. Our approach is in a sense complementary, with the main advantage coming from the fact that the problem is framed into a simple matrix diagonalization form, and that we make use of the Bloch-mode expansion to obtain the exact electric field profile for each mode, which allows us to compute the couplings independently of any approximations. As examples of application, the main results presented in this work concerned radiative effects in the systems of one or two QDs embedded in $Ln$ cavities and the $\mathit{W1}$ waveguide. In the case of one QD, we recover the known results for the Purcell enhancement of the radiative rate and the vacuum Rabi splitting in the strong coupling regime. In the two-QD case, we quantify the strength of the radiative excitation transfer between \textit{spatially separated} QDs, which lies in the $100~\mu\mbox{eV}$ range at short distance. The comparison of the single-mode coupling strength and the energy spacing between modes in $Ln$ cavities of increasing length clearly shows that a crossover occurs -- around $n=150$ for GaAs-based systems -- between single-mode and multi-mode radiative coupling. In the multi-mode case, the radiative coupling strength through each photonic mode is smaller but the overall effective excitation transfer rate still ranges at about $50~\mu\mbox{eV}$, thus suggesting that the $W\mathit{1}$ is an ideal structure for the realization of long-range radiative dot-dot coupling.

These results suggest that the QD-PHC system could be a candidate system to operate as a quantum bus and achieve controlled entangling interaction between distant qubits. This perspective is corroborated by the two following remarks. First, semiconductor QDs have recently seen a tremendous progress\cite{Patton_2005, Press2008, Berezovsky_2008, Greilich_2009, Greilich_2011, Poem_2011, Muller_2012, Godden_2012, Kodriano_2012} towards the physical implementation of qubits that rely on the electron or hole spin as the computational degree of freedom, and on the interband optical transition as the main handle for single-qubit operations. Second, the optical quantum bus technology has already been successfully applied to achieve controlled two-qubit operations in the system of superconducting qubits.\cite{DiCarlo2009} The controlled operation in that case has been achieved by moving in and out of the anti-crossing region in the polariton spectrum arising from radiation-matter coupling. In view of a similar development in the semiconductor QD case, at least three steps are still needed. First, the ability to fabricate site-controlled QDs, in order to position them with respect to the PHC structure. This is nowadays possible thanks to various kinds of growth on a patterned substrate.\cite{Kiravittaya2006,Martin-Sanchez2009,Mehta2007,Mereni2009,Schneider2009} Second, a clear experimental proof of the radiative excitation transfer mechanism at long distance, that might only come from ad-hoc technique such as, for example, the single-QD two-dimensional four-wave-mixing spectroscopy.\cite{Kasprzak_2011} Third, a reliable scheme for dynamically controlling the exciton-photon detuning at sufficiently high speed. For this latter task, extremely promising results are already available on the optical control of the resonant frequency of high-Q cavities, particularly using carrier-induced optical nonlinearity.\cite{Notomi2010,Sato_2012}

An additional challenge is represented by the task of understanding and optimizing disorder effects in the light propagation throughout PHC structures. Apart from small variations of the field profiles\cite{Portalupi_2011}, that should only marginally influence the magnitude of the coupling between one QD and a photon mode, the main effect of disorder is the localization of light in long PHC structures.\cite{Topolancik_2007, Sapienza_2010, Savona_2011, Huisman2012} For the line defects studied in this work in particular, localization is known to compete with the ability of the waveguide to support slow-light propagation,\cite{LeThomas2009, Mazoyer2009} and thus with the enhancement of radiative effects expected in these structures. More specifically, when approaching the band edge in a $W1$ waveguide, the localization length becomes shorter than the decay length related to extrinsic radiation losses. In this limit, there is no more light propagation and the concept of a group velocity of light no longer holds. As an illustrative example of this dramatic effect on light propagation, the statistical fluctuations of the transmission through a waveguide in the light-localization regime increase and the transmission coefficient of a finite-length waveguide takes values uniformly distributed between zero and one, for nominally identical samples and arbitrarily small variations of the frequency.\cite{Mazoyer2010} It is therefore important to accurately characterize the radiative coupling mechanism between distant QDs in a disordered PHC structure. The formalism presented here only relies on the knowledge of the spectrum and field maps of the PHC modes. Together with the possibility to simulate photon modes in very long PHC structures -- offered by the recently developed Bloch-mode expansion method\cite{Savona_2011} -- it therefore represents the election method to carry out a systematic study of disorder effects on the radiative properties of QDs in PHC structures.

\begin{acknowledgments}
This work was supported by the Swiss National Science Foundation through Project No. $200020\_132407$.
\end{acknowledgments}

\bibliography{polaritons_paper}

\end{document}